\documentclass{aa}
\usepackage[varg]{txfonts}
\usepackage{comment}
\usepackage[flushleft]{threeparttable}
\usepackage{graphicx}
\usepackage{color}

\begin{document}

\title{Photoprocessing of formamide ice: route towards prebiotic chemistry in space}
\titlerunning{Photoprocessing of formamide ice}
\author{Maria Angela Corazzi\inst{1,}\inst{2}
  \and Davide Fedele\inst{2}
  \and Giovanni Poggiali\inst{1,}\inst{2}
   \and John Robert Brucato\inst{2}} 
     \authorrunning{Corazzi et al.}

\offprints{M.A.Corazzi,\email{corazzi@arcetri.astro.it}\\
J.R. Brucato,\email{jbrucato@arcetri.astro.it}}
\institute{Departmento of Physics and Astronomy, University of Florence,
Via G. Sansone 1 50019, Sesto Fiorentino (Firenze), Italy
  \and INAF - Astrophysical Observatory of Arcetri,
 Largo E. Fermi 5 50125, Firenze, Italy}
\date{Received .. / Accepted ..}

\abstract {} {Formamide (HCONH$_{2}$) is the simplest molecule containing the peptide bond 
first detected in the gas phase in 
Orion-KL and SgrB$_{2}$. In recent years, it has been observed in 
high temperature regions such as hot corinos, where thermal desorption is responsible for the sublimation of frozen mantles into the gas phase.
The interpretation of observations can benefit from information gathered in the laboratory, where it is possible to simulate the thermal desorption process and to study formamide under simulated space conditions such as UV irradiation.} {Here, two laboratory analyses are reported: we studied formamide photo-stability under UV irradiation when it is adsorbed by space relevant minerals at 63 K and in the vacuum regime. We also investigated temperature programmed desorption of pure formamide ice in the presence of TiO$_{2}$ dust before and after UV irradiation.} {Through these analyses, the effects of UV degradation and the interaction between formamide and different minerals are compared. We find that silicates, both hydrates and anhydrates, offer molecules a higher level of protection from UV degradation than mineral oxides. The desorption temperature found for pure formamide is $220$ K. The desorption temperature increases to $250$ K when the formamide desorbs from the surface of TiO$_{2}$ grains.} {Through the experiments outlined here, it is possible to follow the desorption of formamide and its fragments, simulate the desorption process in star forming regions and hot corinos, and constrain parameters such as the thermal desorption temperature of formamide and its fragments and the binding energies involved.\\
Our results offer support to observational data and improve our understanding of the role of the grain surface in enriching the chemistry in space.} 

\maketitle
\section{Introduction}
Through laboratory analyses, it is possible to study the physical and chemical processes involving prebiotic molecules, the building blocks of life. 
Various different mechanisms have been invoked to explain the formation of interstellar organic complex molecules (iCOMs): grain surface reactions during the cold prestellar period of grain mantle formation \citep{Tielens1982}, ion--molecule reactions in cold and dense gas \citep{Duley1984, Caselli2012}, gaseous phase reactions taking place in dense cores once the icy mantles evaporate in the warm regions \citep{Charnley1995}, and UV- (e.g., \citealt{Ligterink2018}), X-ray- \citep{Ciaravella2019}, and ion- irradiation (e.g., \citealt{Brucato2006}) of icy mantles during the star formation process. In the last three decades, several laboratory studies have focused on the experimental verification of the efficiency of these mechanisms, showing that these processes can
lead to the formation of “complex” organic molecules \citep{munoz2003, urso2017, kanuchova2016, bergantini2014, gerakines2004}. Further studies elucidated the evolution and preservation of these molecules in harsh astronomical environments \citep{Brucato2014, Mennella2015, Baratta2015} where the presence of grains is often invoked. Minerals may play a fundamental role in processes that lead to the emergence of complex molecules because they can act as catalysts, promoting chemical reactions and the synthesis of new molecular species on their surface  or protecting molecules against UV degradation. Moreover, they can adsorb molecules allowing their concentration in the ice mantles \citep{Fornaro2013}.\\ 
Among these latter complex molecules, formamide (HCONH$_{2}$) is of crucial importance.
Chemical reactions of molecules containing H, C, N, and O such as formamide are considered a plausible pathway for synthesis of biomolecules under prebiotic conditions \citep{Oparin1938}. 
Formamide is the simplest molecule containing the peptide bond, which is known to be the basis for assembling proteins and polypeptides starting from amino acids, with a crucial role in the biotic processes of life on Earth. Formamide was first detected in the gaseous phase in two high-mass star forming regions: Orion-KL, an active star forming region, and the giant molecular cloud SgrB$_{2}$ (e.g., \citealt{Turner1991, Nummelin1998, Halfen2011}); 
later it was detected in the comet Hale Bopp \citep{Bockelee-Morvan2000}, whose chemical composition is suspected to be similar to the chemical composition of the primitive Solar Nebula. Furthermore, during landing of Philae aboard Rosetta mission, in situ mass spectrometer data inferred the presence of formamide in the comet nucleus with the highest abundance after water \citep{Goesmann2015}. In recent years, formamide was also observed in two types of low-mass star forming environment: shocked regions by protostellar jets (e.g., \citealt{Codella2017}) and hot corinos (\citealt{Kahane2013}; \citealt{Lopez2015}; \citealt{Marcelino2018}; \citealt{Imai2016}; \citealt{Oya2017}; \citealt{Lee2017}). In high-temperature regions ($>$100 K) such as hot corinos, thermal desorption is responsible for sublimation of frozen mantles into the gas phase. Moreover, outbursting young stars like V883 Ori are good new targets for looking for organic complex molecules that thermally desorb from icy mantles.
 The interpretation of observations can benefit from information coming from the laboratory, where it is possible to simulate the thermal desorption process and UV irradiation of formamide under simulated space conditions. Here we report two laboratory analyses. We focus both on photostability and on thermal desorption of pure formamide, and in the presence of grains, before and after UV irradiation. Section 2 presents our choice of minerals used as analog samples and the experimental setup. In section 3, in situ UV irradiation of pure formamide ice and its adsorption by minerals 
at 63 K are investigated by infrared spectroscopy (FTIR). Section 4 presents  a temperature-programmed desorption (TPD) analysis of pure formamide ice and in the presence of TiO$_{2}$ dust, before and after UV irradiation.

\section{Experimental method}
\subsection{Choice of minerals}
Because silicates are ubiquitous in space, their choice as analog samples is quite straightforward.  
There is evidence of silicates in comets, both from ground observations (e.g.,  \citealt{Shinnaka2018} and \citealt{Ootsubo2019} with 8.2 meter Subaru telescopes or \citealt{Picazzio2019} with SOAR 4.1 meter telescopes) and from space missions (\citealt{Brownlee2006} for Stardust Comet Sample Return Mission and e.g., \citealt{Bockelee2017} for Rosetta Mission). Silicates were also found in the cometary coma of Hale Bopp \citep{Min2005}, the first comet where formamide was detected. Moreover, dust has been observed in circumstellar envelopes around young stars \citep{Cesaroni2017} and in evolved stars and planetary nebulae \citep{Jager1998}. Among silicates, we chose antigorite (Mg,Fe$^{++}$)$_{3}$Si$_{2}$O$_{5}$(OH)$_{4}$ and Forsterite Mg$_{2}$SiO$_{4}$. 
\begin{table}
\caption{\textbf{Minerals and origin}}
\begin{center}
\begin{tabular}{cc}
\hline
Mineral&Origin\\
\hline
Antigorite (MgFe$^{++}$)$_{3}$Si$_{2}$O$_{5}$(OH)$_{4}$&Reichenstein\\
Titanium dioxide TiO$_{2}$&Sintetic\\
Pyrite FeS$_{2}$&Cerro de Pasco\\
Spinel MgAl$_{2}$O$_{4}$&Vesuvio\\
Forsterite Mg$_{2}$SiO$_{4}$&Vesuvio\\
\end{tabular}
\label{tabellaminerali}
\end{center}
\end{table}
Antigorite belongs to the serpentine group; it is a hydrate silicate with a water or hydroxyl group present in the structure and has a lamellar aspect with well-defined cleavages. 
Forsterite Mg$_{2}$SiO$_{4}$ is the magnesium-rich end-member of the olivine group and is one of the most abundant silicate minerals in the Solar System. 
It has been found in meteorites \citep{Weinbruch2000}, cometary dust \citep{Messenger2005}, and a protoplanetary disk \citep{Fujiyoshi2015}, and was observed as tiny crystals in the clouds of gas around a forming star \citep{Poteet2011}. 
Furthermore, Mg-containing minerals are important because the magnesium may have played an important role in prebiotic geochemistry.\\
Furthermore, sulfides were found in the ejecta of comets \citep{Jauhari2008}  and in the cometary coma of Hale Bopp  \citep{Lisse2006}.
Among sulfides, pyrite FeS$_{2}$ was chosen as it is involved in the chemical mechanism responsible for the synthesis of hydrocarbons \citep{Navarro2008}, and, moreover, molecules that contain sulfur are thought to have played an important role in the origin of life and in prebiotic chemistry \citep{Ranjan2017}.\\
Finally, mineral oxides were used such as titanium dioxide TiO$_{2}$ from the rutile group and magnesium alluminium oxide MgAl$_{2}$O$_{4}$ from the spinel group. TiO$_{2}$ comprises a significant amount ($0.63\%$) of the Earth's crust and is preserved as presolar grains extracted from meteorites \citep{Nguyen2018, Nittler2008}. Spinel group minerals are interesting because they have been found in a wide range of terrestrial and extraterrestrial geological environments \citep{Bjarnborg2013, Caplan2017}. 
Table \ref{tabellaminerali} lists the minerals chosen in this work and their origin.
\subsection{Experimental setup}
This work was performed using two different experimental setups. 
The first consists of a vacuum chamber with pressure of $\sim 10^{-5}$ mbar attached to a Bruker VERTEX 70v FTIR infrared spectrometer where formamide samples, adsorbed through the spike technique on various minerals, are irradiated and monitored in-situ with infrared spectroscopy. The interferometer is configured for reflectance spectroscopy and using a combination of optical components (detectors, source, and beamsplitter), it can cover the spectral range from 10 cm$^{-1}$ in the far-infrared (FIR) up to 28000 cm$^{-1}$ in the visible spectral range. 
 In our case, the vacuum chamber was equipped with sapphire windows transparent in the range $8500$ - $1500$ cm$^{-1}$. Praying Mantis Diffuse Reflection Accessory from Harrick was used to allow the acquisition of spectra in reflectance.\\
A second experimental setup was used for TPD analysis. The Hiden Analytical 3F RC 301 Pic Quadrupole Mass Spectrometer (HAL 3F RC) was used for mass spectrometry.
The ion source is an electron impact ionizer with twin-oxide-coated iridium filaments and the detector is a pulse ion counting single channel electron multiplier, with which it is possible to analyze masses from 1 to 300 atomic mass units (a.m.u.). Desorbing surfaces were positioned at 8 mm distance from the ion source. 
A high vacuum ($\sim$ 10$^{-8}$ mbar) chamber with feed-throughs for optical fiber UV irradiation and gas phase deposition was assembled to investigate the thermal desorption process before and after UV irradiation. To simulate the radiation of solar-type stars, a UV-enhanced xenon lamp was used. 
MicrostatHe-R cryostat from Oxford Instruments, cooled using liquid nitrogen (LN$_{2}$), was used in both experimental setups.

\section{Photo-processing}
The formamide sample was purchased from Sigma Aldrich, Merck corporation with a purity $\geq 99.5 \%$. 
We investigated in situ UV irradiation of pure formamide and its adsorption onto minerals at 63 K and at $P\sim 10^{-5}$ mbar.  
LN$_{2}$ was pumped along the transfer line to the cryostat. 
The sample was located in the sample holder, the area and depth of which are $A = 7.07$ mm$^{2}$ and $d=4$ mm, respectively, for a total volume of $V = 0.03$ ml. The sample fixed on the cold finger was inserted into the vacuum chamber interfaced with the interferometer by the reflectance accessory. 
The temperature was controlled with an accuracy of $\pm$0.1 K with \textit{Mercury} iTC by Oxford Instruments. 
The IR reflectance spectra were taken at different irradiation times and the degradation process was followed in real time by observing the changes in spectral features. 
 Each band area, which is proportional to the number of functional groups, was evaluated at different irradiation times. The degradation rate $\beta$ was obtained by fitting the fraction of unaltered molecules, A(t)/A(0), estimated by the band area at a specific irradiation time $t$ normalized on the band area before irradiation, versus time t, using an exponential function:
\begin{equation}
\frac{A(t)}{A_{0}}=B\cdot e^{-\beta\cdot t}+C
\label{degr}
,\end{equation}
where A$_{0}$ is the band area before irradiation, proportional to the initial number of molecules in the sample, $B$ is the fraction of molecules that interacted with UV radiation, $\beta$ is the degradation rate, and $C$ is the fraction of molecules that did not interact with UV radiation because they lie deep in the solid samples. 
In case of band degradation, A(t)/A(0) decreases with increasing irradiation time.
The half-life time t$_{1/2}$ (time necessary to destroy 50$\%$ of the initial molecules) and the UV destruction cross section $\sigma$  were obtained from $\beta$ as follows
\begin{equation}
t_{1/2}=\frac{\ln(2)}{\beta}
\label{tdimezz}
,\end{equation}
\begin{equation}
\beta=\frac{\sigma\cdot \phi_{tot} }{A}
\label{crosssec}
,\end{equation}
where the destruction cross section $\sigma$ is the probability that chemical bonds of formamide are broken by UV radiation. 
Finally, $\phi_{tot}$/A is the incident flux of UV radiation coming out from the optical fiber per unit area integrated with the absorbance of formamide. The photon flux measured from the 
"Consiglio Nazionale delle Ricerche" (CNR) Laboratory of Optics is $2.48\cdot 10^{16}$ photons$\cdot$sec$^{-1}\cdot$cm$^{-2}$. \\
In case of formation of new bands, the kinetic was investigated by the function
\begin{equation}
\frac{A(t)}{A_{max}}=1-e^{-\alpha\cdot t}
\label{form}
,\end{equation}
where A(t) is the band area at the specific time t, which is proportional to the number of molecules formed at the time t; $A_{max}$ is the maximum band area, which is the maximum number of molecules formed; and $\alpha$ is the formation rate. 
In the case of formation of a new band, A(t)/A$_{max}$ increases with increasing irradiation time. 
The formation cross section $\sigma_{f}$ was evaluated as follows
\begin{equation}
\alpha = \frac{\sigma_{f}\cdot \phi_{tot}}{A}
\label{crossform}
,\end{equation}
where the formation cross section $\sigma_{f}$ is the probability that new chemical bonds are formed by UV irradiation.

\begin{figure}
\begin{center}
\resizebox{\hsize}{!}{\includegraphics{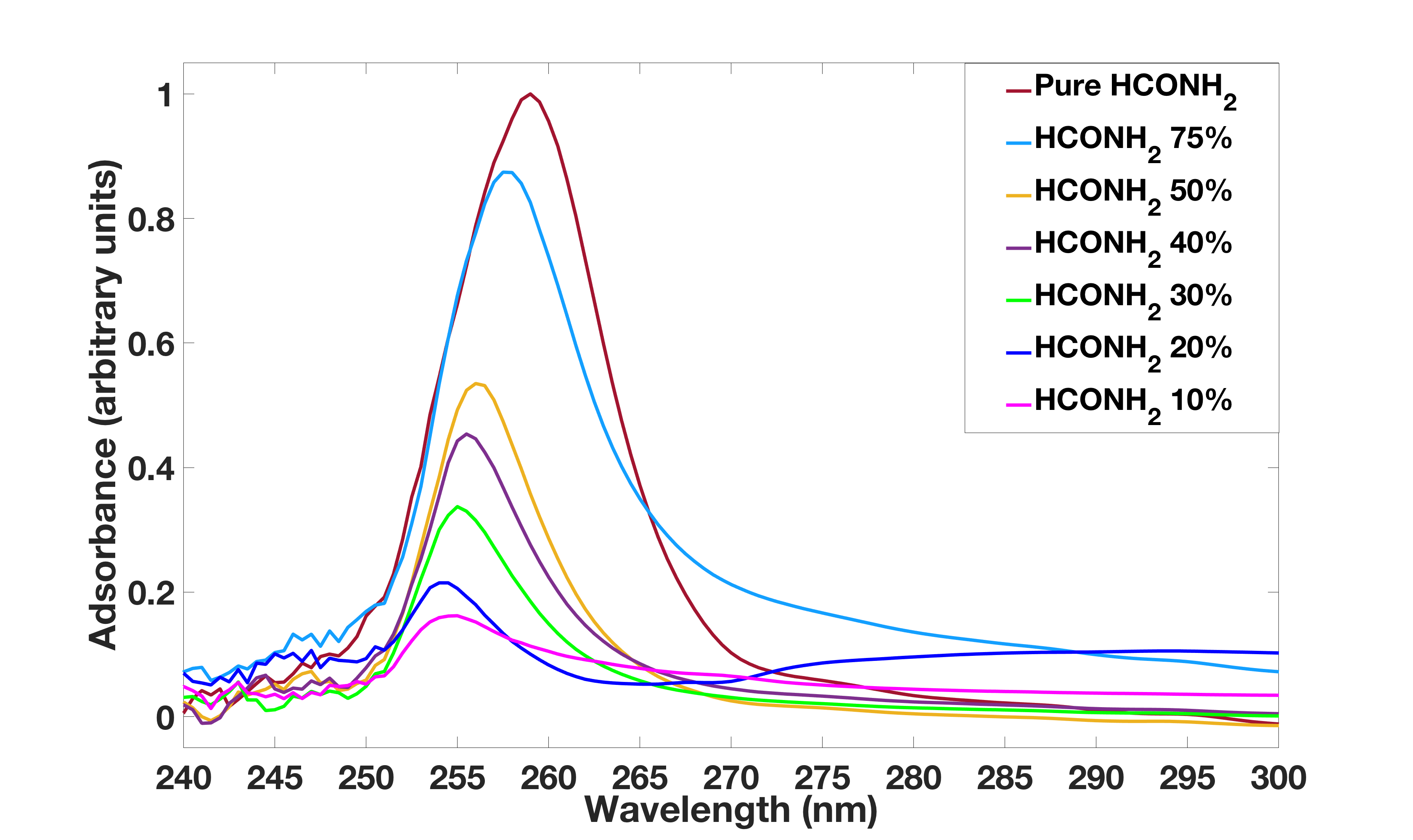}}
\caption{Absorption peaks of solutions with different concentrations of formamide in water.}\label{absorptionpeak}
\end{center}
\end{figure}

\begin{figure}
\begin{center}
\resizebox{\hsize}{!}{\includegraphics{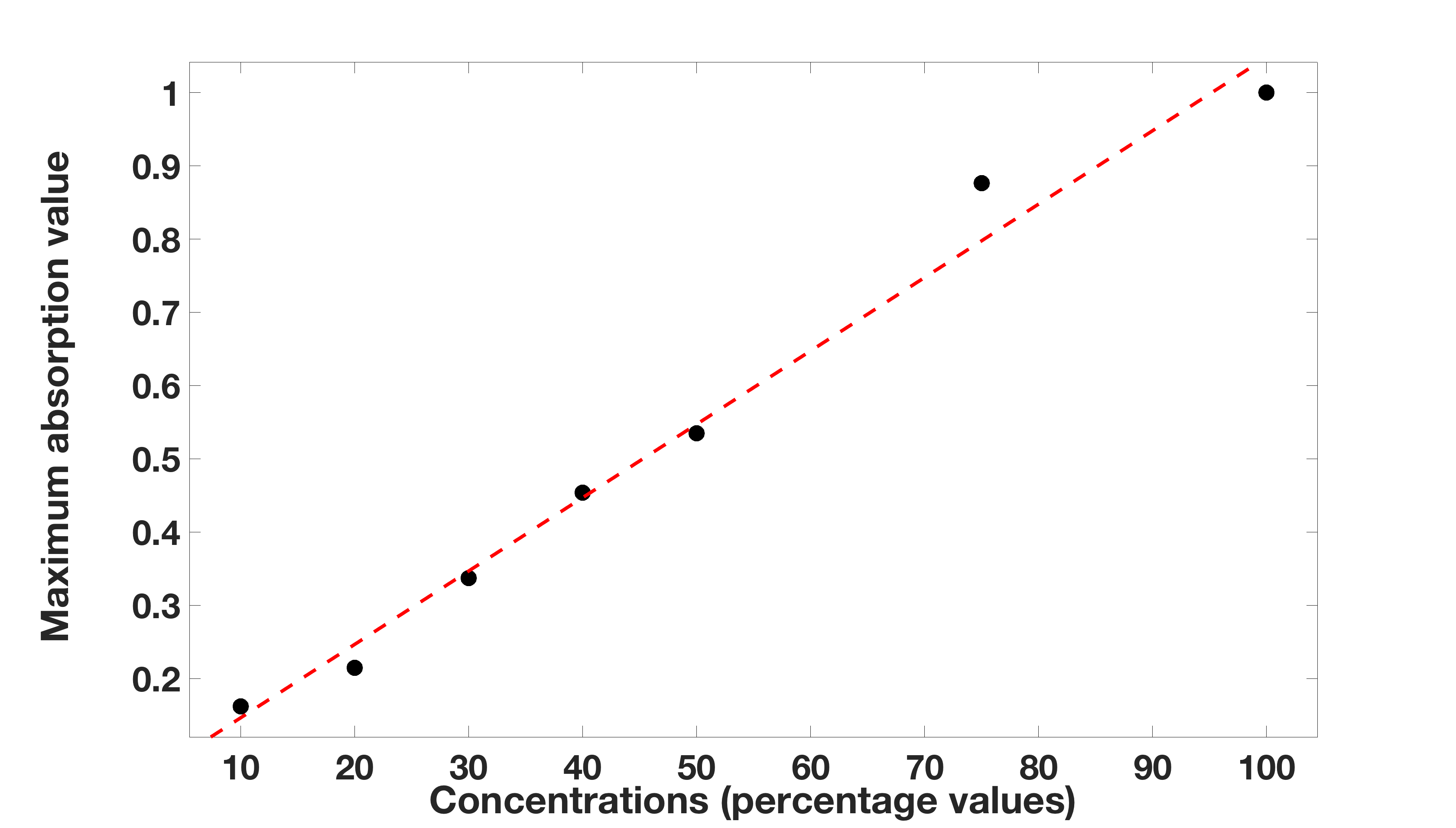}}
\caption{Linear relationship between the maximum absorption value and the formamide concentration.}\label{fitpiccoassorbimento}
\end{center}
\end{figure}

\begin{figure*}
\begin{center}
\includegraphics[scale=0.15]{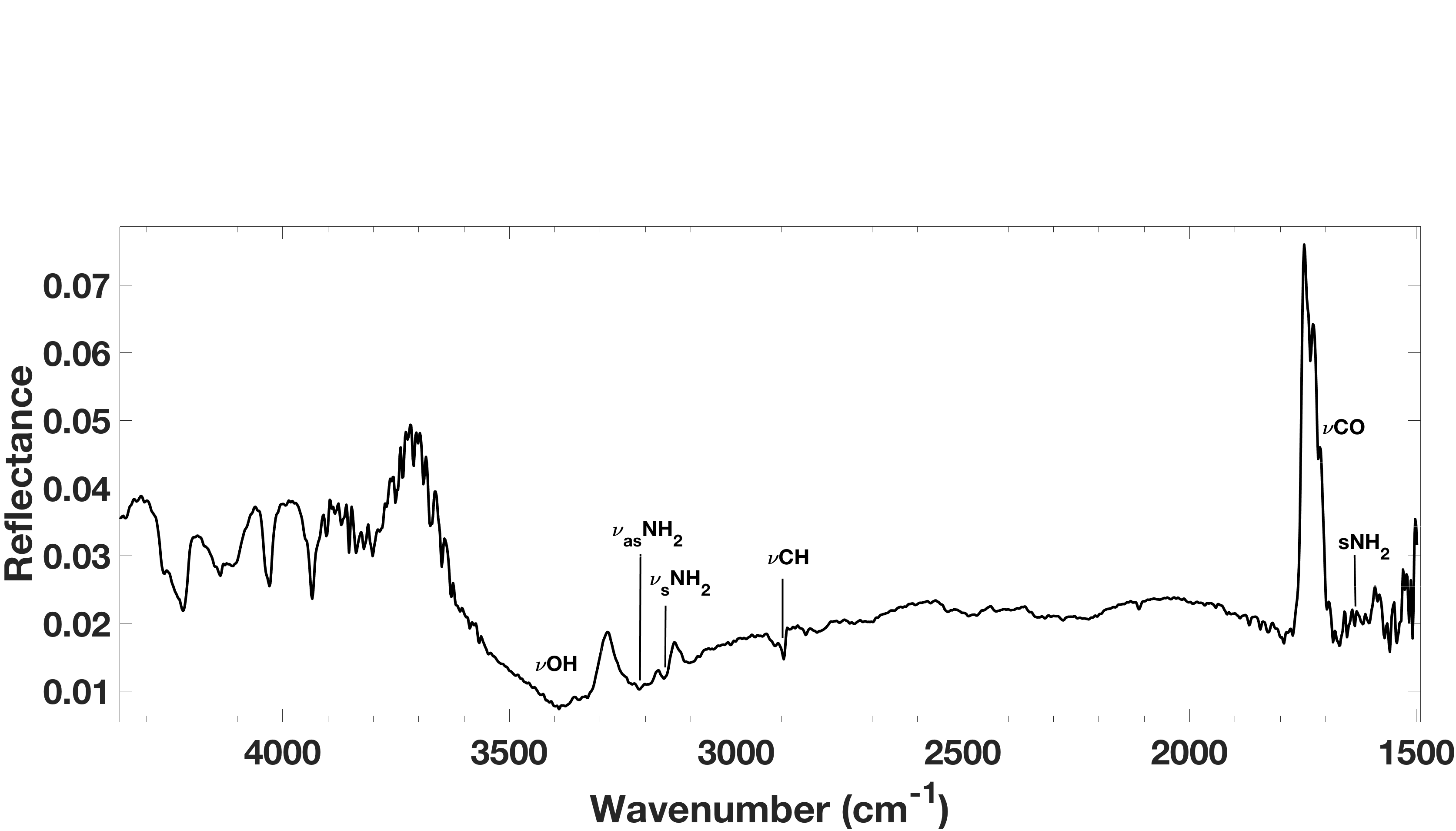}
\caption{Infrared reflectance spectrum of pure formamide ice in the spectral range between 4500 and 1500 cm$^{-1}$ where the main band assignments are present. Vibrational modes: $\nu_{s}$ symmetric stretching; $\nu_{as}$ asymmetric stretching; $\nu$ stretching; $s$ scissoring.
}\label{form4000bande}
\end{center}
\end{figure*}

\begin{figure*}
\begin{center}
\includegraphics[scale=0.15]{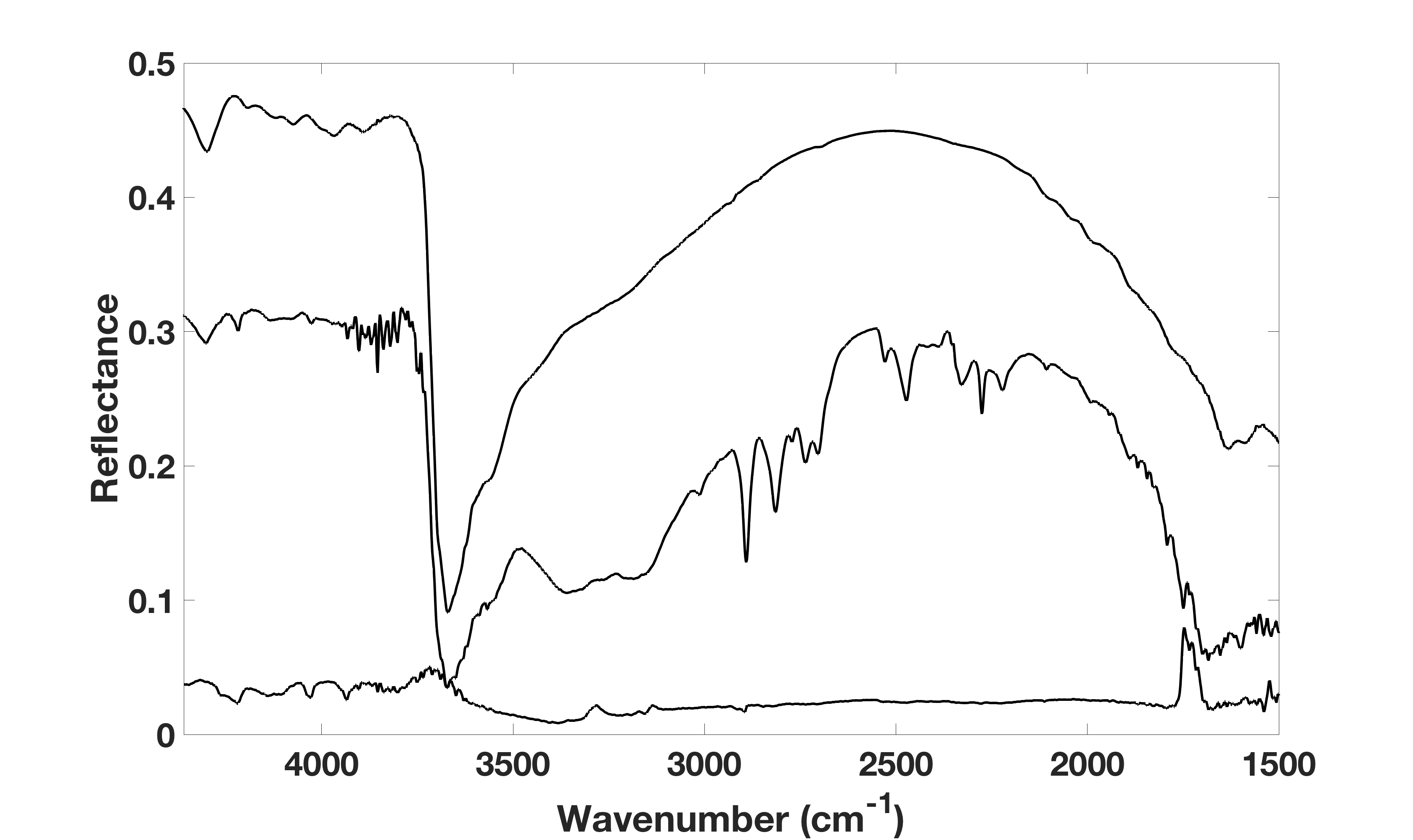}
\caption{From top to bottom: IR reflectance spectrum of antigorite, formamide adsorbed by antigorite, and pure formamide in vacuum. Formamide bands are still present and more intense in the spectrum of formamide adsorbed by the mineral.}\label{serp+form+serp-form}
\end{center}
\end{figure*}

\begin{figure*}
\begin{center}
\includegraphics[scale=0.15]{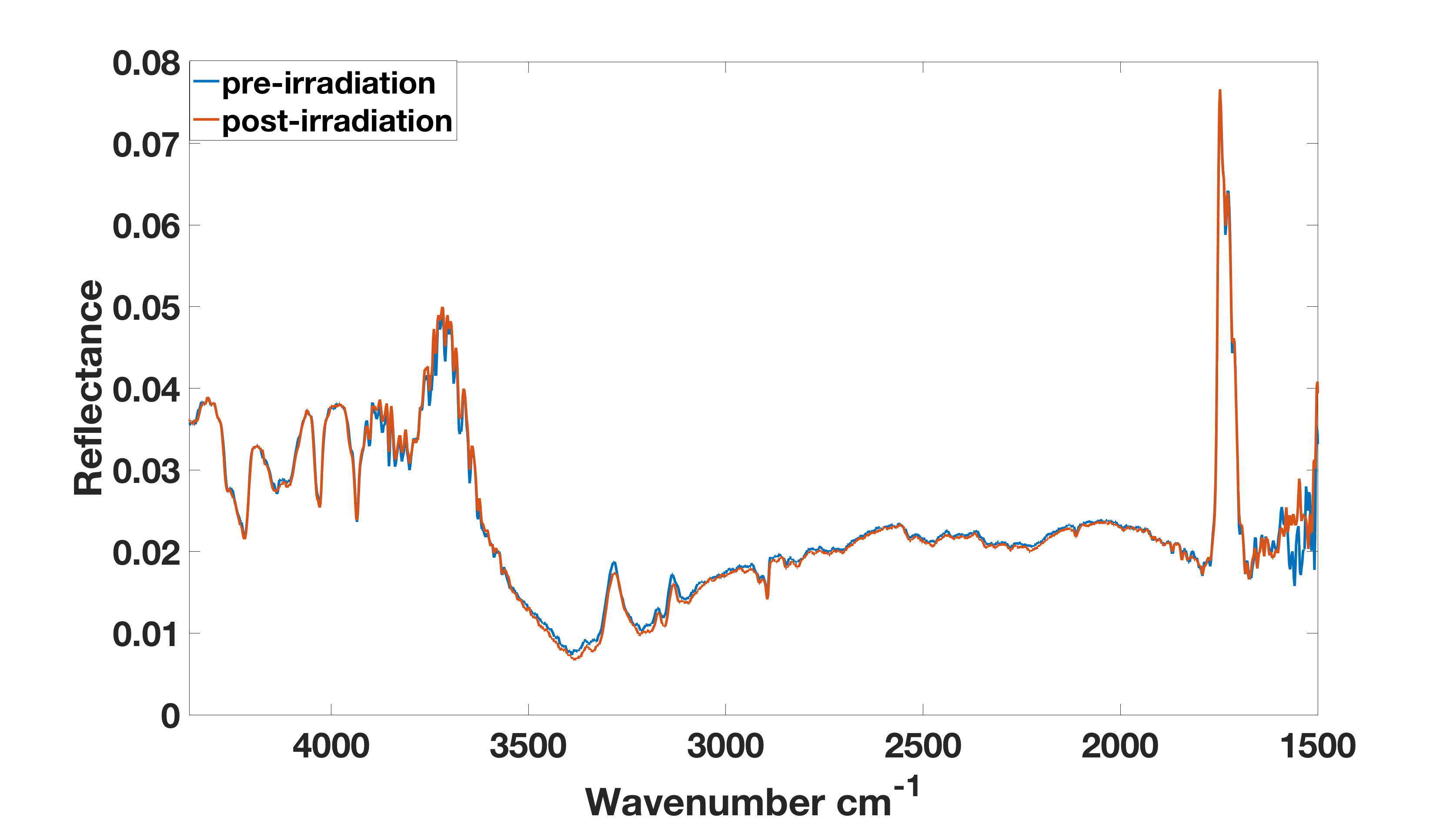}
\caption{
Highlighting the pre- and post-irradiation spectra (total UV irradiation time is  5 hours) of pure formamide ice. The spectrum did not show any significant change.}\label{formamideiceirr}
\end{center}
\end{figure*}

\begin{figure*}
\begin{center}
\includegraphics[scale=0.15]{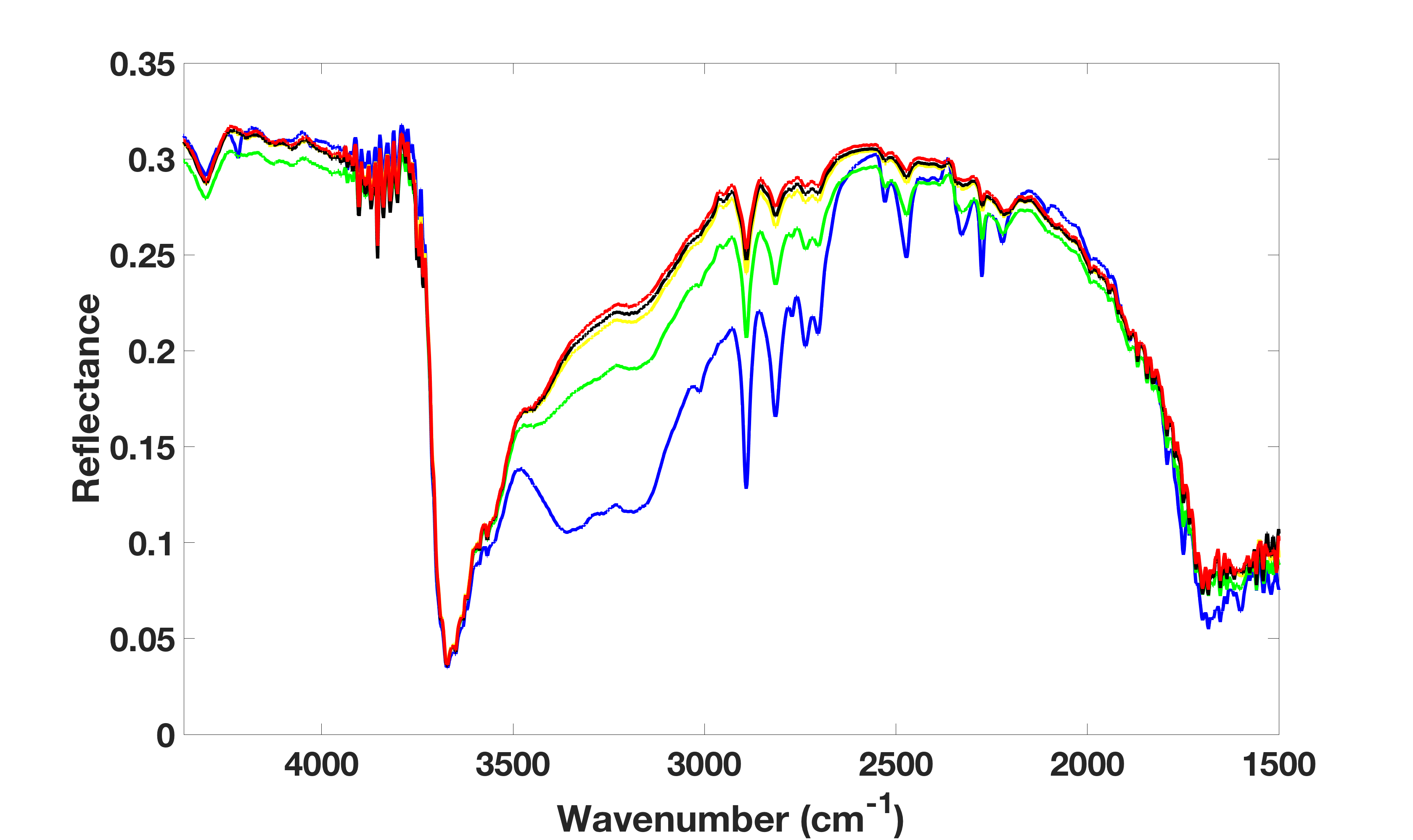}
\caption{Infrared reflectance spectrum of formamide adsorbed on antigorite at $63$ K in vacuum. From bottom to top: before irradiation (blue line), after 1 hour of UV irradiation (green line), after 3 hours of UV irradiation (yellow line), after 4 hours of UV irradiation (black line), after 5 hours of UV irradiation (red line).}\label{serp-form5}
\end{center}
\end{figure*}

\begin{figure*}
\begin{center}
\includegraphics[scale=0.15]{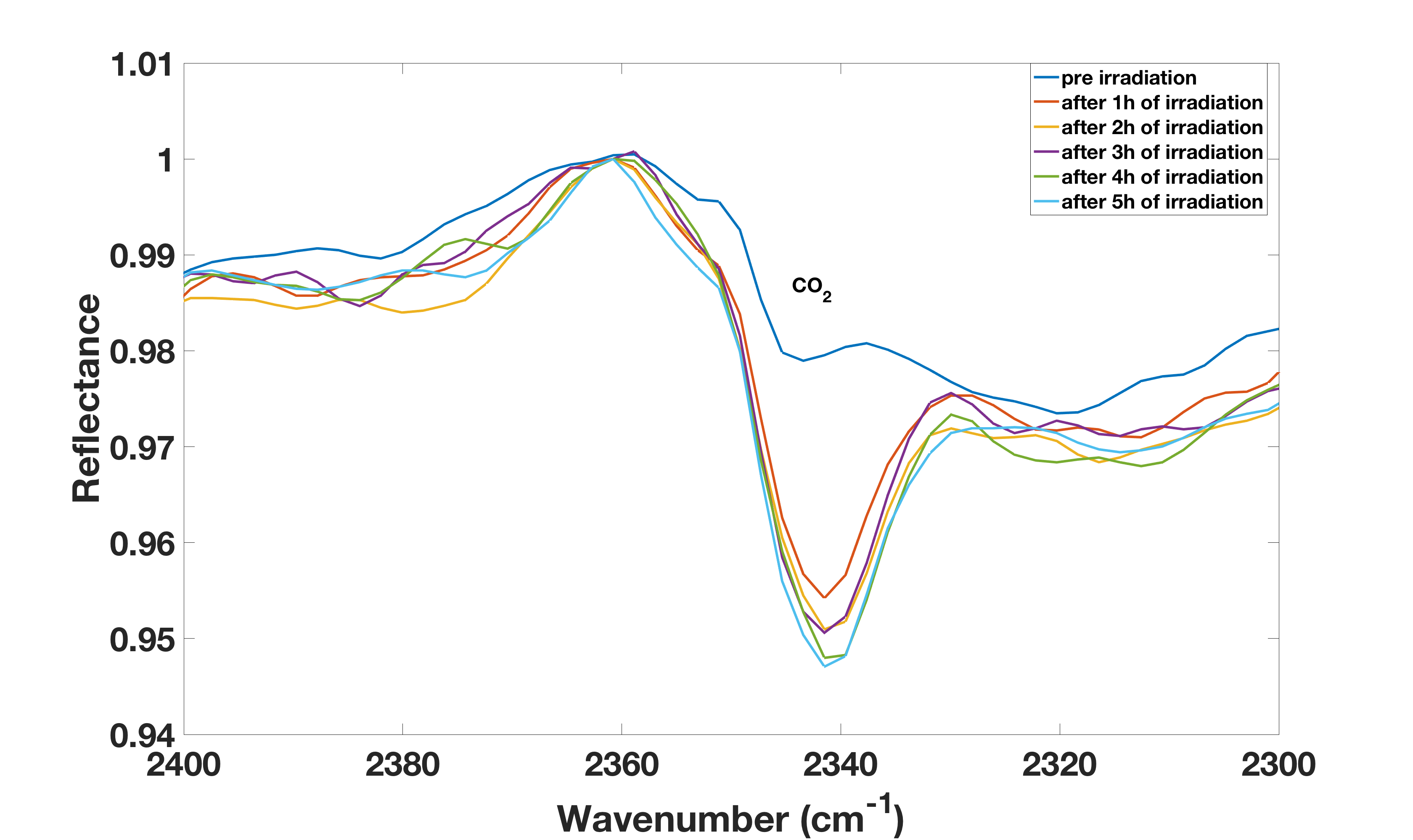}
\caption{Infrared reflectance spectrum of formamide adsorbed by pyrite in the spectral range between 2400 and 2300 cm$^{-1}$. As the irradiation time increased from the higher to the lower spectrum the band at 2341 cm$^{-1}$  due to CO$_{2}$ asymmetric stretching increased in intensity.}\label{piriteformazioneCO2}
\end{center}
\end{figure*}

\paragraph{Results and discussion} 
Initially, we measured the UV absorption of formamide to find out the absorption wavelength range once the samples are subjected to UV irradiation. 
Figure \ref{absorptionpeak} shows the absorption peaks of formamide in water at several concentrations. We found that the absorption is concentrated in the range $200$ - $300$ nm with a maximum between 255 and 260 nm. As the concentration of formamide decreases (from the highest to the lowest curve), the absorption peak shifts towards shorter wavelengths and Figure \ref{fitpiccoassorbimento} shows a linear relationship between the maximum absorption and the concentration of formamide.
\\
The reflectance spectrum of pure formamide at 63 K under simulated space conditions is reported in Figure \ref{form4000bande}, which shows the wavenumber range from 4500 to 1500 cm$^{-1}$ where the main bands analyzed in our work are present. Between 3600 and 3281 cm$^{-1}$, we note the presence of the OH broad band due to residual water present in the chamber and frozen with the sample.
Band assignments are reported in Table \ref{tabassegnamenti}. 
In the first column, we report vibrational modes observed, the second column shows band positions of pure formamide ice compared to the literature positions, which are reported in column three. Band positions of formamide adsorbed on different minerals are also reported in the last columns.\\
\begin{table*}[]
\centering
\begin{center}
\begin{tabular}{cccccccc}
\multicolumn{2}{c}{\textbf{Band assignments}}&&&&&&\\
\\
Mode&Band (cm$^{-1}$)&&\multicolumn{5}{c}{Formamide adsorbed on}\\
&Formamide&Literature&Antigorite&TiO$_{2}$&Pyrite&Spinel&Forsterite\\
$\nu_{as}$ NH$_{2}$&3266&$3368^{1}$&3350&3298&...&3285&3275\\
$\nu_{s}$ NH$_{2}$&3162&$3181^{1}$&3166&3166&...&3166&3166\\
$\nu$ CH&2895&$2891^{1}$&2891&2891&2890&2891&2891\\
...&...&...&2815&2815&2816&2816&2815\\
$\nu$ CH&...&$2770-2763^{2}$&2770&...&...&2772&2771\\
$\nu$ CO&...&...&2473&2471&2473&2475&2474\\
$\nu_{as}$ CO$_{2}$&...&...&...&...&...&2348&...\\
$\nu$ CO&...&...&2329&2328&...&2323&2328\\
$\nu_{as}$ NCO&...&...&2275&2275&2276&2277&2277\\
$\nu$ CO&...&$1754^{3}$&1750&1750&1750&1750&1750\\
$\nu$ CO&1715&$1713^{2}$&1716&1717&1717&1716&1717\\
$\nu$ CO&1697&$1698-1704^{1}$&1699&1699&1699&1699&1699\\
$\nu$ CO&1684&$1685-1681^{1}$&1684&1684&1684&1684&1684\\
$s$ NH$_{2}$&1635&$1631^{4,1}$&1635&1635&1635&1635&1635\\
\end{tabular}
\caption{First column: vibrational modes; second column: bands observed in the IR reflectance spectrum of pure formamide during laboratory work; third column: band assignments based on published experimental and theoretical studies: (1) \citealt{Sivaraman2013}, (2) \citealt{Nguyen1986}, (3) \citealt{McNaughton1999}, (4) \citealt{Brucato2006}; other columns: bands observed in the IR reflectance spectrum of formamide adsorbed by antigorite, TiO$_{2}$, pyrite, spinel, forsterite. Vibrational modes: $\nu_{s}$ symmetric stretching; $\nu_{as}$ asymmetric stretching; $\nu$ stretching; $s$ shissoring}
\label{tabassegnamenti}
\end{center}
\end{table*}
Table \ref{tabassegnamenti} shows that the bands of pure formamide are still present when formamide is adsorbed by minerals, even if a shift of peak positions occurs. An example of how spectral features change before and after adsorption onto minerals is reported in Figure \ref{serp+form+serp-form}, where the IR reflectance spectrum of pure formamide, antigorite mineral, and formamide adsorbed by antigorite in a vacuum at 63 K are shown. The spectrum of pure formamide ice shows two neighboring bands at 3266 and 3162 cm$^{-1}$ due respectively to NH$_{2}$ asymmetric and symmetric stretching and a band at 2895 cm$^{-1}$ due to CH stretching \citep{Sivaraman2013}. These bands are still present and more intense in the spectrum of formamide adsorbed by antigorite, but they are shifted to the wavenumbers 3350, 3166, and 2891 cm$^{-1}$ respectively (Table \ref{tabassegnamenti}). The spectra of formamide adsorbed on the other minerals also show these bands.
 When formamide is adsorbed by TiO$_{2}$, spinel, and forsterite, the band at 3266 cm$^{-1}$ is shifted to the wavenumbers 3298, 3285, and 3275 cm$^{-1}$ respectively, while the 3162 cm$^{-1}$ band is present at 3166 cm$^{-1}$ in the presence of all minerals. On the contrary, pyrite does not show these two neighboring bands. In the presence of TiO$_{2}$, spinel, and forsterite, the 2895 cm$^{-1}$ formamide band is shifted to the wavenumber 2891 cm$^{-1}$, the same shift is obtained in the presence of antigorite. In the case of pyrite, it is moved to 2890 cm$^{-1}$. At lower wavenumbers, the formamide spectrum shows bands due to CO stretching: 1715 cm$^{-1}$ \citep{Nguyen1986}, 1697 cm$^{-1}$, and 1684 cm$^{-1}$ \citep{Sivaraman2013}. All these bands are still present in the mineral spectra. The band of this functional group at 1715 cm$^{-1}$ shows a small shift to 1716 cm$^{-1}$ when formamide is adsorbed by antigorite and spinel and to 1717 cm$^{-1}$ in the presence of TiO$_{2}$, pyrite, and forsterite. The band at 1697 cm$^{-1}$ shifts to 1699 cm$^{-1}$ in all the other spectra.  Alternatively, the band at 1684 cm$^{-1}$ has the same position in all spectra. The 1635 cm$^{-1}$ formamide band due to NH$_{2}$ scissoring (\citealt{Brucato2006}; \citealt{Sivaraman2013}) also shows no shift and it is present in the same position in all spectra (see Table \ref{tabassegnamenti}).
\\
Now we present our study of the effects of UV photolysis, first irradiating pure formamide and then when it is adsorbed by minerals.  
Pure formamide ice did not show any significant degradation (i.e., IR band intensity does not decrease) even after 5 hours of UV irradiation as shown in Figure \ref{formamideiceirr}. 
This result suggests that formamide is highly stable upon radiation under the conditions simulated in our experiment.  
However, finding pure formamide in the solid state in space is unlikely. It is more plausible that formamide is formed in a mixture of other iCOMs as an icy mantle on dust grains. When formamide is adsorbed by minerals and subjected to the same irradiation experiments (equal irradiation time with same photon flux, therefore an equal amount of UV energy), stability is compromised.
Figure \ref{serp-form5} shows how the spectral features of formamide adsorbed by antigorite changed as the irradiation time increased. 
From this figure,  after 5 hours of radiation 
the intensity of the bands at $3350$ and $3166$ cm$^{-1}$ due to NH$_{2}$ stretching can immidiately be seen to be notably reduced, that is, the radiation has degraded the molecule, preferentially breaking the amino group bonds.   
We observed a reduction of intensity also for the band at 2891 cm$^{-1}$ which is due to CH stretching and those at 2473, 2329, and 1750 cm$^{-1}$ which is due to CO stretching. 
For these bands, we evaluated from the kinetic analysis of the degradation process a cross section $\sigma$ of $\sim 10^{-20}$cm$^{2}$. Similar results were obtained for formamide adsorbed by forsterite, which unlike antigorite is an anhydrate silicate. With forsterite, the two bands due to NH$_{2}$ stretching showed no change after 5 hours of irradiation, while the  intensity of the bands at 2771 and 2474 cm$^{-1}$ due to  CH and CO stretching, respectively, was reduced. Compared to antigorite, for irradiating formamide adsorbed by forsterite we observed intensity changes in a smaller number of bands, but the degradation process occurred with the same cross section of $10^{-20}$cm$^{2}$. When formamide was adsorbed by TiO$_{2}$, the bands that showed variation during the  irradiation experiment were the two bands at 3298 and 3166 cm$^{-1}$ due to NH$_{2}$ stretching and the band at 2891 cm$^{-1}$ due to CH stretching. With the spinel, on the other hand, the bands at 2323 and 2348 cm$^{-1}$ due to CO stretching and CO$_{2}$ asymmetric stretching, respectively, showed changes. When formamide was adsorbed both by TiO$_{2}$ and by spinel, the cross sections found were $10^{-19}$cm$^{2}$, that is, an order of magnitude greater than that obtained for the silicates. From these results, we can divide minerals according to their effect on photodegradation of formamide. 
Under the experimental conditions with which the photo-stability of formamide was studied, we found that the silicates are more effective in protecting formamide against UV degradation compared with mineral oxides, the difference being of an order of magnitude. \\
When formamide was adsorbed by pyrite, the degradation of the bands was not observed even after 5 hours of UV irradiation, and therefore pyrite is capable of protecting the molecule from UV degradation. Interestingly, only when formamide was adsorbed by pyrite or forsterite did we observe the formation of a new band at 2341 cm$^{-1}$ due to CO$_{2}$ asymmetric stretching. Therefore, during the irradiation, CO$_{2}$ formation occurred within the icy matrix (Figure \ref{piriteformazioneCO2}). 
Tables 3, 4, and 5 report our kinetic analysis of the degradation and formation process performed for all bands of each sample.
\begin{table*}[]
\begin{center}
\begin{tabular}{cccccc}
\multicolumn{4}{c}{\textbf{Photodissociation parameters in simulated space conditions}}&&\\
\\
\multicolumn{6}{c}{IR reflectance spectrum of formamide adsorbed by antigorite at $63$ $K$ in vacuum}\\
Peak (cm$^{-1}$)&Mode&$\beta$ (min$^{-1}$)&B&C&$R^{2}$\\ \hline
3350-3166&$\nu_{as,s}$ NH$_{2}$&$0,029 \pm 0,005$&$0,87 \pm 0,05$&$0,05 \pm 0,04$&0,98\\
2891&$\nu$ CH&$0,023 \pm 0,005$&$0,46 \pm 0,06$&$0,51 \pm 0,05$&0,98\\
2473&$\nu$ CO&$0,026 \pm 0,008$&$0,75 \pm 0,07$&$0,24 \pm 0,05$&0,99\\
2329&$\nu$ CO&$0,023 \pm 0,004$&$0,84 \pm 0,05$&$0,20 \pm 0,04$&0,97\\
1750&$\nu$ CO&$0,018 \pm 0,009$&$0,6 \pm 0,1$&$0,38 \pm 0,08$&0,97\\
&&&&&\\
\multicolumn{6}{c}{IR reflectance spectrum of formamide adsorbed by TiO$_{2}$ at $63$ K in vacuum}\\
Peak (cm$^{-1}$)&Mode&$\beta$ (min$^{-1}$)&B&C&$R^{2}$\\ \hline
3298-3166&$\nu_{as,s}$ NH$_{2}$&$0,17 \pm 0,07$&$0,20 \pm 0,02$&$0,776 \pm 0,011$&0,9\\
2891&$\nu$ CH&$0,08 \pm 0,02$&$0,127 \pm 0,013$&$0,847 \pm 0,007$&0,94\\
&&&&&\\
\multicolumn{6}{c}{IR reflectance spectrum of formamide adsorbed by spinel at $63$ K in vacuum}\\
Peak (cm$^{-1}$)&Mode&$\beta$ (min$^{-1}$)&B&C&$R^{2}$\\ \hline
2348&$\nu_{as}$ CO$_{2}$&$0,305 \pm 0,001$&$0,994 \pm 0,001$&$0,006 \pm 0,001$&1\\
2323&$\nu$ CO&$0,556 \pm 0,001$&$0,510 \pm 0,001$&$0,490 \pm 0,001$&1\\
&&&&&\\
\multicolumn{6}{c}{IR reflectance spectrum of formamide adsorbed by forsterite at $63$ K in vacuum}\\
Peak (cm$^{-1}$)&Mode& $\beta$ (min$^{-1}$)&B&C&$R^{2}$\\
\hline
2815&...&$0,015 \pm 0,010$&$0,054 \pm 0,013$&$0,954 \pm 0,010$&0,98\\
2771&$\nu$ CH&$0,02 \pm 0,02$&$0,15 \pm 0,12$&$0,85 \pm 0,16$&0,8\\
2474&$\nu$ CO&$0,015 \pm 0,008$&$0,040 \pm 0,006$&$0,952 \pm 0,005$&0,88\\
\end{tabular}
\caption{First column: bands that showed variations during the  irradiation experiment; second column: vibrational modes; third column: the degradation rate $\beta$ obtained through equation \ref{degr}; fourth and fifth columns: fraction of molecules that interacted and did not interact with UV radiation respectively (with boundary condition B+C=1); sixth column: $R^{2}$ points to the goodness of fit with which $\beta$ was obtained.}
\label{tabfotodissparam}
\end{center}
\end{table*}
\begin{table*}[]
\begin{center}
\begin{tabular}{cccc}
\multicolumn{2}{c}{\textbf{Cross section and half-life times }}&\\
\\
\multicolumn{4}{c}{Formamide adsorbed by antigorite}\\
Peak (cm$^{-1}$)&Mode&$\sigma$ (cm$^{2}$)&$t_{1/2}$ (min)\\ \hline
3350-3166&$\nu_{as,s}$ NH$_{2}$&$(1,9 \pm 0,6)\cdot 10^{-20}$&$24 \pm 4$\\
2891&$\nu$ CH&$(1,5 \pm 0,9)\cdot 10^{-20}$&$30 \pm 13$\\
2473&$\nu$ CO&$(1,7 \pm 0,8)\cdot 10^{-20}$&$27 \pm 8$\\
2329&$\nu$ CO&$(1,5 \pm 0,5)\cdot 10^{-20}$&$30 \pm 5$\\
1750&$\nu$ CO&$(1,2 \pm 0,8)\cdot 10^{-20}$&$39 \pm 19$\\
&&&\\
\multicolumn{4}{c}{Formamide adsorbed by TiO$_{2}$}\\
Peak (cm$^{-1}$)&Mode&$\sigma$ (cm$^{2}$)&$t_{1/2}$ (min)\\ \hline
3298-3166&$\nu_{as,s}$ NH$_{2}$&$(1,14 \pm 0,64)\cdot 10^{-19}$&$4 \pm 2$\\
2891&$\nu$ CH&$(0,5 \pm 0,2)\cdot 10^{-19}$&$9 \pm 2$\\
&&&\\
\multicolumn{4}{c}{Formamide adsorbed by spinel}\\
Peak (cm$^{-1}$)&Mode&$\sigma$ (cm$^{2}$)&$t_{1/2}$ (min)\\ \hline
2348&$\nu_{as}$ CO$_{2}$&$(2,1 \pm 0,3)\cdot 10^{-19}$&$2,273 \pm 0,007$\\
2323&$\nu$ CO&$(3,7 \pm 0,6)\cdot 10^{-19}$&$1,2467 \pm 0,0002$\\
&&&\\
\multicolumn{4}{c}{Formamide adsorbed by forsterite}\\
Peak (cm$^{-1}$)&Mode&$\sigma$ (cm$^{2}$)&$t_{1/2}$ (min)\\ \hline
2815&...&$(1,0 \pm 0,8)\cdot 10^{-20}$&$46 \pm 31$\\
2771&$\nu$ CH&$(1 \pm 2)\cdot 10^{-20}$&$35 \pm 35$\\
2474&$\nu$ CO&$(1,0 \pm 0,7)\cdot 10^{-20}$&$46 \pm 25$\\
\end{tabular}
\caption{For bands reported in Table \ref{tabfotodissparam} that showed variations during the  irradiation experiment. The cross section $\sigma$ (third column) and half-lifetime t$_{1/2}$ (fourth column) were evaluated through the equations \ref{crosssec} and \ref{tdimezz} respectively.}
\label{tabcrosssec}
\end{center}
\end{table*}
\begin{table*}[]
\begin{center}
\begin{tabular}{ccccc}
\multicolumn{2}{c}{\textbf{Parameters and cross section for band-formation process}}&\\
\\
\multicolumn{5}{c}{Formamide adsorbed by pyrite}\\
Peak (cm$^{-1}$)&Mode&$\alpha$ (min$^{-1}$)&$R^{2}$&$\sigma_{f}$ (cm$^{2}$)\\ \hline
2341&$\nu_{as}$ CO$_{2}$&$0,043 \pm 0,011$&0,8&$(3 \pm 1)\cdot 10^{-20}$\\
&&&\\
\multicolumn{5}{c}{Formamide adsorbed by forsterite}\\
Peak (cm$^{-1}$)&Mode&$\alpha$ (min$^{-1}$)&$R^{2}$&$\sigma_{f}$ (cm$^{2}$)\\ \hline
2340&$\nu_{as}$ CO$_{2}$&$0,0121 \pm 0,0016$&0,94&$(0,8 \pm 0,2)\cdot 10^{-20}$\\
\end{tabular}
\caption{First column: bands that increased in intensity during the irradiation experiment; second column: vibrational modes; third column: the formation rate $\alpha$ obtained through the equation \ref{form}; fourth column: $R^{2}$ points to the goodness of fit with which $\alpha$ was obtained; fifth column: formation cross section evaluated through equation \ref{crossform}.}
\label{tabcrossform}
\end{center}
\end{table*}

\section{Thermal desorption}
In the high vacuum chamber, HVC ($5\cdot 10^{-8}$ mbar), formamide vapor was deposited on the cold finger at 63 K through a valve system. 
The molecules condensed and formed an icy film which was subjected to in situ UV irradiation. After irradiation, the sample was heated at a constant rate of 0.6 K$\cdot$sec$^{-1}$. As the cold finger warmed up, the condensed molecules desorbed, entered the mass spectrometer, and were detected; TPD curves were obtained for selected masses. 
We used the same experimental setup to study the thermal desorption of formamide condensed onto dust grains. The grains were placed on the sample holder of the cryostat and inserted into the high vacuum chamber. In this case, when formamide entered the chamber, it condensed onto the surface of the grains. \\ 
Usually, a TPD curve is described by the Polanyi-Wigner equation (e.g., \citealt{attard1998}):
\begin{equation}
\frac{-d\theta}{dT}=\frac{A}{h_{r}}\cdot \theta^{m}\cdot e^{-E_{d}/RT}
\label{PW}
,\end{equation}
where
\begin{itemize}
\item $\frac{-d\theta}{dT}$ is the desorption rate,
\item $\theta$  the surface coverage,
\item $m$  the desorption order,
\item $A$  the pre exponential factor in sec$^{-1}$,
\item $h_{r}$  the heating rate,
\item $E_{d}$  the desorption energy, and
\item $R$  the ideal gas constant.
\end{itemize}
The shape of the TPD curve is the convolution of two terms, the Arrhenius equation $\frac{A}{h_{r}}\cdot e^{-E_{d}/RT}$ that increases exponentially with the temperature and the surface coverage $\theta^{m}$. The surface coverage is a function that decreases as the temperature increases due to molecular desorption. The convolution of the two terms determines the desorption temperature and then the desorption energy $E_{d}$.

\begin{figure}
\begin{center}
\resizebox{\hsize}{!}{\includegraphics{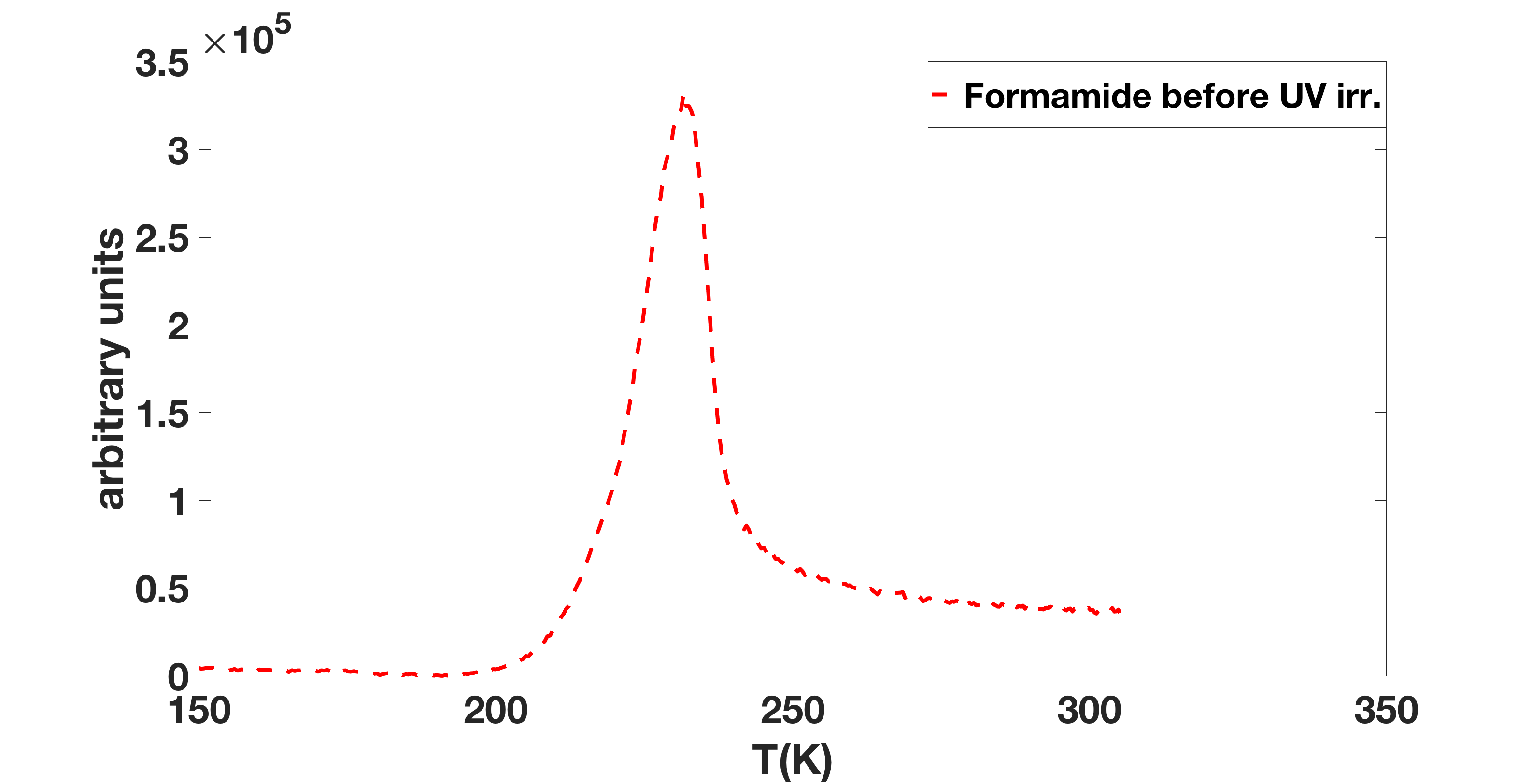}}
\caption{Temperature-programmed
desorption curve of pure formamide before UV irradiation.}\label{TPDform}
\end{center}
\end{figure}

\begin{figure*}
\begin{center}
\includegraphics[scale=0.15]{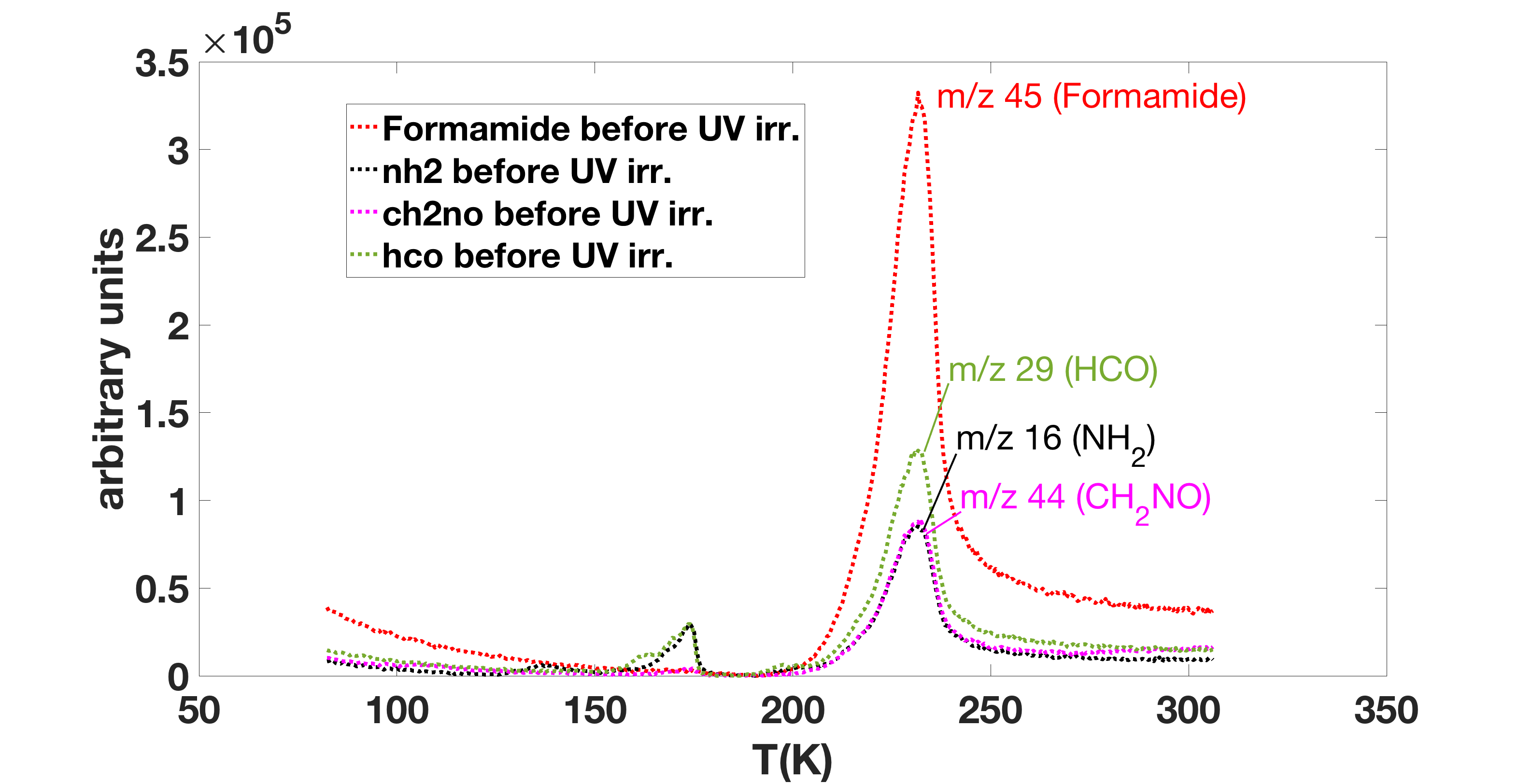}
\caption{Temperature-programmed
desorption curves of formamide (red dashed line - m/z 45), NH$_{2}$ (black dashed line - m/z 16), CH$_{2}$NO (purple dashed line - m/z 44) and HCO (green dashed line - m/z 29) before UV irradiation.}\label{TPDnoirr}
\end{center}
\end{figure*}

\begin{figure*}
\begin{center}
\includegraphics[scale=0.15]{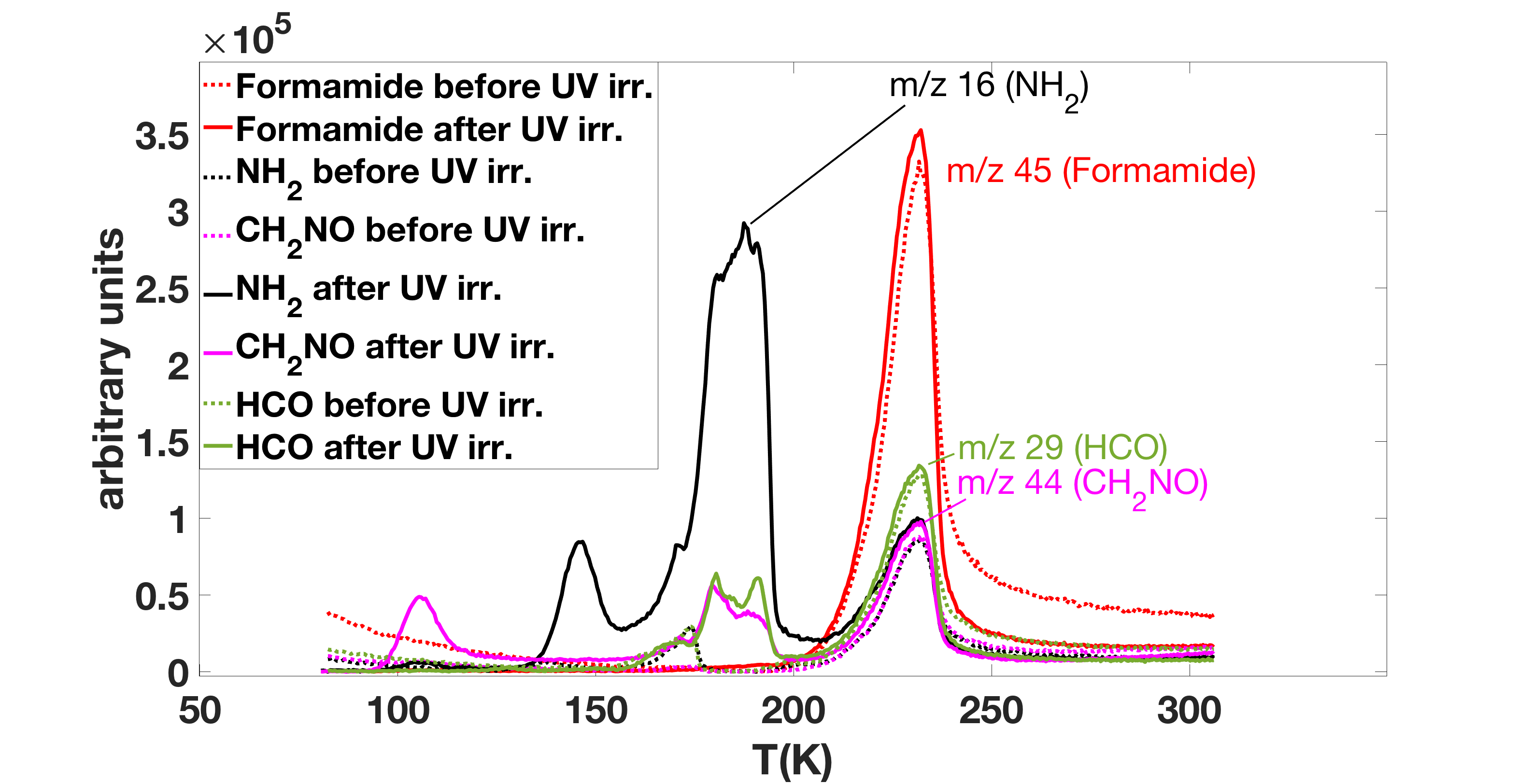}
\caption{Temperature-programmed
desorption curves of formamide (red line - m/z 45), NH$_{2}$ (black line - m/z 16), CH$_{2}$NO (purple line - m/z 44) and HCO (green line - m/z 29) before (dashed line) and after (continuous line) five hours of UV irradiation.}\label{form5oredirr}
\end{center}
\end{figure*}

\begin{figure}
\begin{center}
\resizebox{\hsize}{!}{\includegraphics{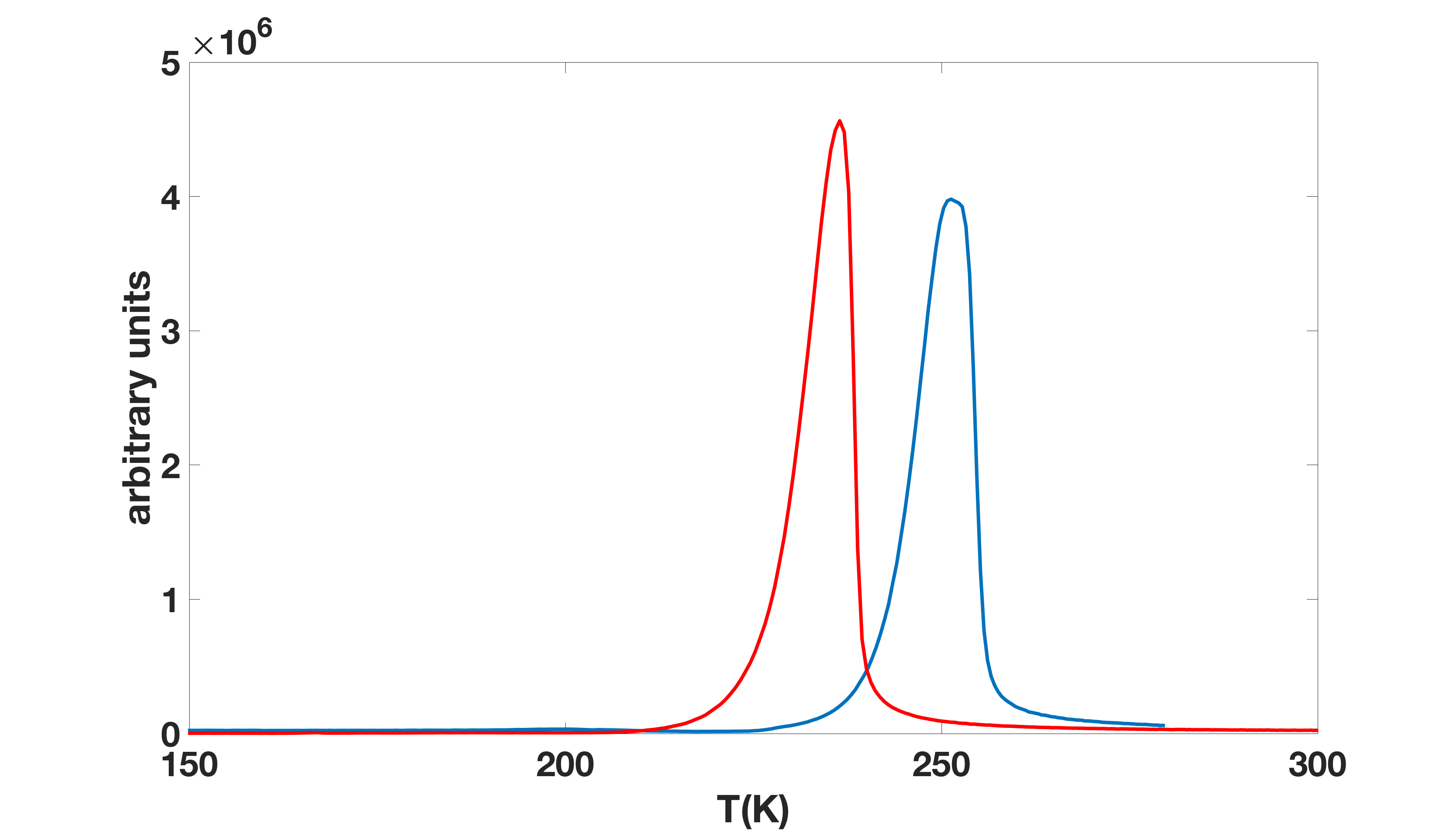}}
\caption{Red line: TPD curve of pure formamide condensed on the cold finger of cryostat; Blue line: TPD curve of formamide condensed on TiO$_{2}$.}\label{confrontiform}
\end{center}
\end{figure}

\begin{figure*}
 \begin{center}
 \includegraphics[scale=0.34]{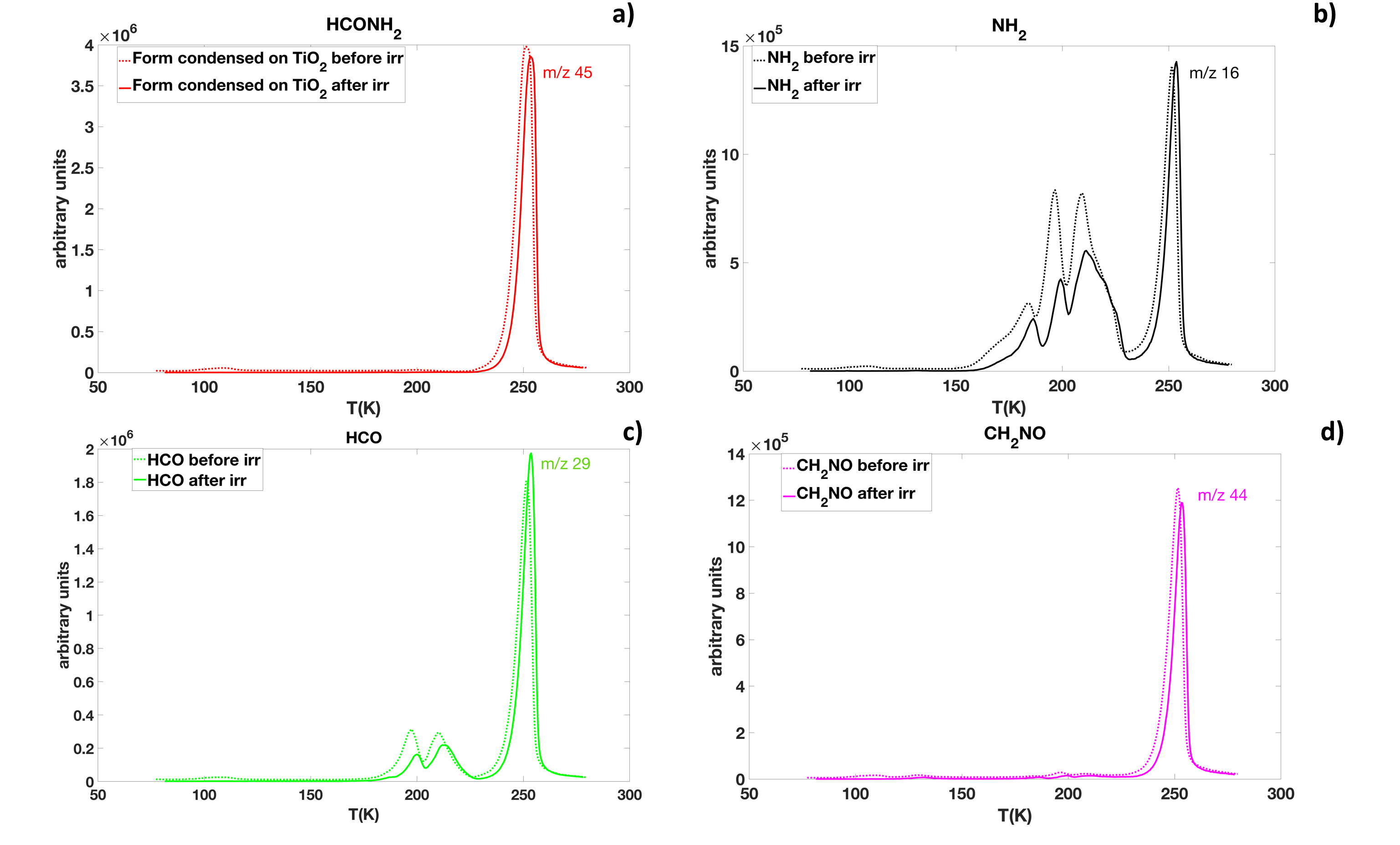}
 \caption{ (a) TPD curves of formamide ( m/z 45) condensed on TiO$_{2}$ before (dashed lines) and after (continuous lines) 4 hours UV irradiation; (b), (c), and (d): TPD curves of major fragments (NH$_{2}$, HCO and CH$_{2}$NO respectively) before (dashed lines) and after (continuous lines) 4 hours of UV radiation.}\label{formeframmentiTiO2beforeandafter}
 \end{center}
 \end{figure*}

\paragraph{Results and discussion}
Figure \ref{TPDform} shows the TPD curve of pure formamide ice at 45 a.m.u. before UV irradiation. The maximum desorption peak was found at $\sim 220$ K.
During the same experiment, it was found that also the signals at 16 a.m.u. (NH$_{2}$), 29 a.m.u. (HCO), and 44 a.m.u. (CH$_{2}$NO) increased at 220 K, the same temperature of sublimation of formamide (Figure \ref{TPDnoirr}). These signals were attributed to formamide fragments due to dissociation by the mass spectrometer.
Figure \ref{form5oredirr} shows the TPD curves of formamide (red line),  NH$_{2}$ (black line),  CH$_{2}$NO (purple line), and HCO (green line) before (dashed line) and after (continuous line) five hours of UV irradiation.
We noticed that after five hours of irradiation, at $\sim 180$ K the counts of NH$_{2}$ (black continuous line) reached the value of $3\cdot 10^{5}$ arbitrary units (a.u.) and the counts of CH$_{2}$NO (purple continuous line) and of HCO (green continuous line) became $5\cdot 10^{4}$ a.u. 
At this temperature, formamide was not yet desorbed, but it was still condensed on the cold finger of the cryostat. 
Therefore, these fragments were not due to the dissociation of formamide inside the mass spectrometer, but they were photoproducts due to UV irradiation.  
In our experiment, the desorption temperature of water occurred at 180 K.
The presence of water was due to residual deposition in the high vacuum chamber. It is reasonable to think, therefore, that the sublimation of water ice was responsible for releasing more volatile species. This result is probably a close representation of what can happen in space. The capability of water to trap molecules with lower sublimation temperatures is a well-known process occurring in space and it can influence the gas-phase composition of the  interstellar medium \citep{Aznar2010}. 
Figure \ref{form5oredirr} also shows that, after UV irradiation, there was an increase in NH$_{2}$ counts at $150$ K (maximum value of $1\cdot 10^{5}$ a.u.). Also, this peak is connected to water. At $150$ K, there is the conversion of cubic crystalline ice to hexagonal ice \citep{Collings2004}, which is responsible for the desorption of more volatile species trapped in the water ice structure. From these results, it seems that desorption of NH$_{2}$, HCO, and CH$_{2}$NO was driven by the transition of water ice structure. The same phenomenon probably occurs in the space where water is the most abundant molecule.
The TPD curve of NH$_{2}$ after irradiation shows a small increase in the counts even at a lower temperature $\sim 100$ K; this desorption feature is also present in the TPD curve of CH$_{2}$NO where it reached the value of $5\cdot 10^{4}$ a.u. (Figure \ref{form5oredirr}). We found that CH$_{2}$NO in the UHV regime desorbed at $\sim 100$ K. Once desorbed, it split into its fragments for dissociation by mass spectrometry, and therefore the desorption feature of NH$_{2}$ at $100$ K is due to the fragmentation of CH$_{2}$NO inside the mass spectrometer.
\\
Going forward with our investigation, we reproduced the condensation, irradiation, and desorption experiments with a substrate of TiO$_{2}$ dust. Figure \ref{confrontiform} shows the differences in TPD curves between pure formamide and in the presence of TiO$_{2}$ grains. We observed evidence for a change in desorption temperature between the two experiments. Formamide desorption from TiO$_{2}$ dust occurred at higher temperatures, that is, around 30 K above the temperature at which desorption takes place for pure formamide. A higher desorption temperature is direct evidence of the interaction described by the Van der Waals forces that were occurring between the molecule and the grains. The molecule interacts and diffuses into the grains and this is confirmed by the values of the binding energy.  
When formamide desorbed directly from the cold finger of the cryostat (copper chromate surface), the binding energy found was $(7.5 \pm 0.7)\cdot 10^{3}$K; while when it desorbed from TiO$_{2}$ dust, the binding energy found was higher $(1.18 \pm 0.07)\cdot 10^{4}$K, as reported in the second column of Table \ref{tablefit}.\\
Temperature-programmed
desorption curves of formamide and fragments condensed on TiO$_{2}$ before and after UV radiation are shown in Figure \ref{formeframmentiTiO2beforeandafter}.
Interestingly, in presence of TiO$_{2}$, all the TPD curves shifted towards higher temperatures after the irradiation experiment. 
Before UV radiation, formamide condensed on the surface of TiO$_{2}$ desorbed at $250.8$ K (red dashed line Figure \ref{formeframmentiTiO2beforeandafter} a); after UV radiation it desorbed at $253.2$ K (red continuous line Figure \ref{formeframmentiTiO2beforeandafter} a). Temperature-programmed
desorption curves for major fragments reported in Fig. \ref{formeframmentiTiO2beforeandafter} (b, c, and d) showed the same shift to the right. 
Table \ref{tablefit} shows the values found for desorption energies before and after UV irradiation, evaluated with equation \ref{PW}. 
\begin{table*}
\begin{center}
\caption{\textbf{Best-fit values of TPD curves}}
\begin{tabular}{l|l|l}
&Before UV&After UV\\
\hline
Pure formamide&$E_{d}/R=(7.5 \pm 0.7)\cdot 10^{3}$K&$E_{d}/R=(5.9 \pm 0.3)\cdot 10^{3}$K\\
Formamide on TiO$_{2}$&$E_{d}/R=(1.18 \pm 0.07)\cdot 10^{4}$K&$E_{d}/R=(1.35 \pm 0.08)\cdot 10^{4}$K\\
\hline
\end{tabular}
\label{tablefit}
\end{center}
\end{table*}

\section{Astrophysical context and the importance of temperature-programmed desorption analyzes}
In recent years, formamide has been observed in low-mass star formation environments such as shocked regions by protostellar jets (e.g., \citealt{Codella2017}).
Protostellar shocks produced by episodic jets of matter propagate at supersonic velocity through dense gas surrounding the star. In these shocked regions, the presence of a large number of molecules is due to mechanical release by gas--grain collisions and grain--grain collisions (\citealt{Lopez2019}; \citealt{Codella2017}).\\
Aside from protostellar shocks, the other low-mass star formation environments where formamide has been detected are hot corinos (\citealt{Kahane2013}; \citealt{Lopez2015}; \citealt{Marcelino2018}; \citealt{Imai2016}; \citealt{Oya2017}; \citealt{Lee2017}).
Hot corinos are the  inner, compact ($<$ 100 astronomical units), hot ($>$100 K) regions of some Class 0 protostars (\citealt{Ceccarelli2004}; \citealt{Ceccarelli2007}; \citealt{Lopez2019}). Interestingly, formamide has only been detected in protostars that host a hot corino (\citealt{Sakai2013}; \citealt{Lopez2019}). This evidence implies that, besides the release in protostellar shocks, the presence of formamide is strictly linked to hot (> 100K) regions where 
thermal desorption is the responsible process for sublimation of frozen mantles into the gas phase. 
This is in very good agreement with our TPD analysis which requires a temperature of 220 K for thermal desorption of formamide (Figure \ref{TPDform}). 
The desorption energies published by \cite{Wakelam2017}, who proposed a new model to compute the binding energy of species to water ice surfaces, are adopted in the chemical model of \cite{Quenard2018}, which shows that formamide needs environments with temperatures of greater than 200 K to be present in the gas phase,  otherwise it remains frozen on dust grains. 
However, our experiments show that these energies, consistent with those we found for formamide desorbed directly from the cold finger, are greater when considering adsorption on grain surfaces: using TiO$_{2}$ grains, the binding energies increased (Table \ref{tablefit}) and the sublimation temperature was 30 K higher than that found for pure formamide (Figure \ref{confrontiform}). Therefore, in the chemical models of sublimation, it is essential to take into account physisorption of iCOMs on grain surfaces and their diffusion in order to  correctly describe the desorption process, that is, to constrain desorption temperatures and binding energies.
Following \cite{Collings2004}, the desorption temperature ranges from 220 K at a 0.6 K sec$^{-1}$ heating rate as measured in the laboratory, to 137 K at 1 K century$^{-1}$, the typical heating rate of hot cores.
 Furthermore, our experiments show something more than the desorption temperature and the binding energy of formamide. The molecular fragments observed in the gas phase can be used to indirectly measure the presence of formamide: we found that during the desorption experiment, NH$_{2}$ counts reached the value of $3\cdot 10^{5}$ a.u. at 180 K, six times higher than HCO and CH$_{2}$NO counts (Figure \ref{form5oredirr}), and during the TiO$_{2}$ grain desorption experiment, NH$_{2}$ counts reached the value of $8\cdot 10^{5}$ a.u. at 200 K, four times higher than HCO counts (Figure \ref{formeframmentiTiO2beforeandafter}). Therefore, these ratios between the abundance of NH$_{2}$ and the abundance of HCO and CH$_{2}$NO (e.g., [NH$_{2}$]/[HCO]$\sim$4) can be an indicator of the presence of formamide.

\section{Conclusion and future perspective} 
Here, we study the effects of UV degradation and the thermal desorption process of pure formamide and formamide condensed on certain minerals found in space. In this way, we simulated the formamide that is initially adsorbed by the  grain surface within regions of star formation before desorption at  high temperatures and entry into the gas phase. 
To study the interaction with different minerals and to try to reproduce the process occurring in hot star forming regions,  two analyses  were performed in the laboratory in parallel: a UV photodissociation analysis and a TPD analysis.\\
Through the UV photodissociation analysis, we studied first the effect of UV radiation on pure formamide ice and then the role that the different minerals have in its photodegradation.
From our experiments, silicates were found to offer molecules a higher level of protection   from UV degradation than mineral oxides which accelerated the degradation process by an order of magnitude. 
These results can be used to predict the fate and the average life of formamide by extrapolating the expected UV radiation in hot regions of star formation or circumstellar environments of young stars.\\
In the second analysis, the thermal desorption of formamide was simulated. Thermal desorption is a physical process that is certainly present in warm regions ($>$100 K) such as hot corinos where formamide has been 
observed several times; for example, formamide has been detected in IRAS 16293-2422 \citep{Kahane2013}, in NGC 1333 IRAS 4A \citep{Lopez2015}, in B335 \citep{Imai2016}, in L483 \citep{Oya2017}, and in HH 212 \citep{Lee2017}.\\ 
To date, formamide has not been detected in protoplanetary disks, where, for a solar-type star, the region where the temperature reaches the values for the desorption and release into the gaseous phase of water and COMs (water snow line) is very close to the star: less than 5 astronomical units (e.g., \citealt{Cieza2016}). 
This region is difficult to solve because of its small size and in future extremely high-sensitivity observations will be required.\\
A new perspective is provided by objects such as the FU Ori disk in which the young central star undergoes a sudden increase in brightness which leads to heating of the disk and quick expansion of the snow lines to large radii. This phenomenon has been observed by  \cite{Cieza2016} in the protoplanetary disk around the protostar V883 Ori ($1.3$ $M_{\odot}$).
These latter  authors observed that the increase in brightness extended the water snow line up to 42 au. Methanol has been detected in this disk by \cite{vanthoff2018}: the transitions observed suggest the thermal desorption of methanol from dust grains from the surface layers beyond or within the water snow line, which would therefore have extended up to 100 au. 
Thanks to the increase in temperature of the disk around V883 Ori and the consequent thermal desorption of the molecules, five iCOMs were detected: methanol, acetone, acetonitrile, acetaldehyde, and methyl formate \citep{Lee2019}.
Complex molecules are difficult to observe in protoplanetary disks, but outbursting young stars like V883 Ori are good targets to look for iCOMs that thermally desorb from icy mantles \citep{vanthoff2018}. Formamide has not yet been revealed, but the next step for the scientific community is surely to search for formamide in this kind of object. 
\\
Laboratory studies on thermal desorption are fundamental to constrain parameters such as the thermal desorption temperature of a given molecule and its fragments, and the binding energies involved.
All these laboratory studies are in support of the interpretation of  formamide observations in star forming regions, hot corinos, and the certain forthcoming observations of formamide in objects like the disk of FU Ori. At the same time, laboratory analysis of the photo-stability of molecules is important to assess the fate of formamide and the role of the grain surface in driving prebiotic chemistry in space.\\

\paragraph{Aknowledgements}
The authors are grateful to the anonymous referee for very helpful suggestions. We wish to thank the Italian Space Agency for co-funding the Life in Space project (ASI N. 2019-3-U.0).\\
This work was supported by (i) the PRIN-INAF 2016 "The Cradle of Life - GENESIS-SKA (General Conditions in Early Planetary Systems for the rise of life with SKA)", and (ii) the program PRIN-MIUR 2015 STARS in the CAOS - Simulation Tools for Astrochemical Reactivity and Spectroscopy in the Cyberinfrastructure for Astrochemical Organic Species (2015F59J3R, MIUR Ministero dell'Istruzione, dell'Università della Ricerca e della Scuola Normale Superiore).\\
D.F. acknowledges financial support provided by the Italian Ministry of Education, Universities and Research, project SIR (RBSI14ZRHR). \\
Thanks also to Lara Fossi for her contribution and support during laboratory experiments.
\bibliography{Bibliografia}
\bibliographystyle{aa}
\end{document}